\documentclass[12pt, twocolumn, reprint]{iopart}
\usepackage{comment}
\usepackage{xcolor}
\usepackage{array}
\usepackage{arydshln}
\usepackage{amsmath}
\usepackage{subcaption}
\usepackage{gensymb}
\usepackage{cancel}
\usepackage{lipsum}
\usepackage{multirow}
\usepackage[hashEnumerators,smartEllipses]{markdown}

\usepackage[backend=biber,style=chem-angew,maxnames=4,urldate=comp,url=false]{biblatex}
\usepackage{float}
\usepackage{subcaption}
\usepackage{color,soul}
\usepackage{setspace}
\usepackage{import}
\usepackage{titlesec}
\usepackage[export]{adjustbox}
\usepackage{tocloft}
\usepackage{supertabular}

\newcommand{\sub}[2]{#1_\mathrm{#2}}
\usepackage{amsmath,physics,amssymb,slashed}
\usepackage{blkarray}
\usepackage{graphicx} %
\graphicspath{{./Images/}} %
\usepackage{caption} %

\usepackage{longtable} %

\usepackage[normalem]{ulem}
\useunder{\uline}{\ul}{}
\emergencystretch=\maxdimen
\usepackage{tikz}
\usetikzlibrary{calc,trees,positioning,arrows.meta,chains,shapes.geometric,  decorations.pathreplacing,decorations.pathmorphing,shapes,  matrix,shapes.symbols}
\tikzset{>={Latex[width=2mm, length=1.2mm]}}

\usepackage{float}

\usepackage{listings}
\usepackage{xcolor}

\usepackage{textcomp}

\definecolor{codegreen}{rgb}{0,0.6,0}
\definecolor{codegray}{rgb}{0.5,0.5,0.5}
\definecolor{codepurple}{rgb}{0.58,0,0.82}
\definecolor{backcolour}{rgb}{0.95,0.95,0.92}

\lstdefinestyle{mystyle}{
    backgroundcolor=\color{backcolour},   
    commentstyle=\color{codegreen},
    keywordstyle=\color{magenta},
    numberstyle=\tiny\color{codegray},
    stringstyle=\color{codepurple},
    basicstyle=\ttfamily\footnotesize,
    breakatwhitespace=false,         
    breaklines=true,                 
    captionpos=b,                    
    keepspaces=true,                 
    numbers=left,                    
    numbersep=5pt,                  
    showspaces=false,                
    showstringspaces=false,
    showtabs=false,                  
    tabsize=2,
}
\usepackage{pdfpages}

\usepackage{dirtytalk}
\usepackage[margin=.6in,includefoot,footskip=30pt]{geometry}
\usepackage{makecell}
\usepackage{graphicx,booktabs}
\usepackage{subcaption}
\newcommand{\labelphantom}[1]{%
{\phantomsubcaption%
\label{#1}}%
}%
\makeatletter
\renewcommand\p@subfigure{\thefigure-}
\makeatother

\begin{document}

\twocolumn[
  \begin{@twocolumnfalse}
    \title{Characterizing the negative triangularity reactor core operating space with integrated modeling}
\author{%
H.~S.~Wilson$^{1}$,
A.~O.~Nelson$^{1}$,
J. McClenaghan$^{2}$,
P. Rodriguez-Fernandez$^{3}$
J. Parisi$^{4}$
and
C.~Paz-Soldan$^{1}$
}

\address{$^{1}$~Department of Applied Physics and Applied Mathematics, Columbia University, New York, NY 10027, USA}
\address{$^{2}$~General Atomics, San Diego, CA 92121, USA}
\address{$^{3}$~Plasma Science and Fusion Center, Massachusetts Institute of Technology, Cambridge, MA 02139, USA}
\address{$^{4}$~Princeton Plasma Physics Laboratory, Princeton, NJ 08540, USA}

\maketitle
\begin{abstract}
Negative triangularity (NT) has received renewed interest as a fusion reactor regime due to its beneficial power-handling properties, including low scrape-off layer power and a larger divertor wetted area that facilitates simple divertor integration. NT experiments have also demonstrated core performance on par with positive triangularity (PT) H-mode without edge-localized modes (ELMs), encouraging further study of an NT reactor core. In this work, we use integrated modeling to scope the operating space around two NT reactor strategies. The first is the high-field, compact fusion pilot plant concept MANTA (The MANTA Collaboration \textit{et al} 2024 \textit{Plasma Phys. Control. Fusion} \textbf{66} 105006) and the second is a low field, high aspect ratio concept based on work by Medvedev \textit{et al} (Medvedev \textit{et al} 2015 \textit{Nucl. Fusion} \textbf{55} 063013). By integrating equilibrium, core transport, and edge ballooning instability models, we establish a range of operating points with less than 50 MW scrape-off layer power and fusion power comparable to positive triangularity (PT) H-mode reactor concepts. Heating and seeded impurities are leveraged to accomplish the same fusion performance and scrape-off layer exhaust power for various pressure edge boundary conditions. Scans over these pressure edge conditions accommodate any current uncertainty of the properties of the NT edge and show that the performance of an NT reactor will be extremely dependent on the edge pressure. The high-field case is found to enable lower scrape-off layer power because it is capable of reaching high fusion powers at a relatively compact size, which allows increased separatrix density without exceeding the Greenwald density limit. Adjustments in NT shaping exhibit small changes in fusion power, with an increase in fusion power density seen at weaker NT. Infinite-$n$ ballooning instability models indicate that an NT reactor core can reach fusion powers comparable to leading PT H-mode reactor concepts while remaining ballooning-stable. Seeded krypton is leveraged to further lower scrape-off layer power since NT does not have a requirement to remain in H-mode while still maintaining high confinement. We contextualize the NT reactor operating space by comparing to popular PT H-mode reactor concepts, and find that NT exhibits competitive ELM-free performance with these concepts for a variety of edge conditions while maintaining relatively low scrape-off layer power.
\end{abstract}
  \end{@twocolumnfalse}
]

\section{Introduction}
\label{sec:intro}
While great progress has been made in the fusion energy field toward power-plant relevant plasma performance and divertor technology, a major outstanding challenge remains: the coupling of a high-performance core to a realistic exhaust solution. This challenge has shifted the focus of some plasma core modeling efforts away from maximizing power output and toward optimizing power handling potential. Fusion pilot plant (FPP) tokamak concepts have been dominated by positive triangularity (PT) plasmas operating in high confinement mode (H-mode). H-mode is accessed when a PT plasma is given sufficient heating and fueling and is characterized by the formation of an edge transport barrier with high pressure gradients called a pedestal \cite{Burrell_1994}. Given its higher confinement, H-mode is generally seen as a desirable regime for an FPP. However, the power through the scrape-off-layer $\sub{P}{SOL}$ must be above the L-H mode transition power $\sub{P}{LH}$ estimated by scaling laws \cite{Martin_2008} to sustain H-mode. In a PT H-mode reactor-class device, this leads to heat fluxes that will be difficult for plasma facing components to sustain without an advanced divertor \cite{Gunn_2017}. Even if we assume we can advance divertor and material technology to sustain reactor-level H-mode heat loads, H-mode bears yet another challenge: it is accompanied by edge localized modes (ELMs) \cite{HILL1997182}. ELMs are instabilities that can result in large energy fluences to plasma facing materials if not mitigated in some way \cite{Gunn_2017}. As power, current, and magnetic field are increased to reactor-relevant levels, the machine damage from ELMs is likely to be intolerable \cite{EICH2017, Gunn_2017}.

There are multiple regimes that exhibit higher confinement than the traditional low-confinement mode (L-mode) with smaller ELMs than H-mode or that avoid them all together. These include I-mode, QH-mode, EDA H-mode, and quasi-continuous exhaust (QCE), among others \cite{Paz-Soldan_2021, VIEZZER2023}. While PT no-ELM or small-ELM regimes are better than PT H-mode for device longevity, they often have sensitive access conditions depending on the heating and fueling scheme or do not yet provide sufficient fusion performance improvement over L-mode to support a realistic FPP design \cite{Paz-Soldan_2021}. 

Recently, the negative triangularity (NT) regime has resurfaced as a potential solution to the core-edge integration challenge in a reactor-class tokamak \cite{Marinoni_2021, Kikuchi_2019}. The ``upper" and ``lower" triangularities of a toroidal plasma are defined as $\sub{\delta}{u,l} = (\sub{R}{geo}-\sub{R}{u,l})/a$, where $\sub{R}{geo}$ is the geometric major radius, $\sub{R}{u}$ is the major radius at the highest point of the separatrix, $\sub{R}{l}$ is the major radius at the lowest point of the separatrix, and $a$ is the plasma minor radius. The average triangularity $\delta$ is the mean of the upper and lower triangularities. Because NT plasmas have $\sub{R}{u,l}>\sub{R}{geo}$, NT x-points are located at a larger major radius than PT x-points. This allows for more space for a divertor and a larger divertor-wetted area, both of which are beneficial from a power-handling perspective \cite{Medvedev_2015, Kikuchi_2019}. Further, the NT regime has been shown to be ELM-free as long as the triangularity is sufficiently negative, even in cases where $\sub{P}{SOL}$ exceeds $\sub{P}{LH}$ by a significant margin \cite{Nelson_2023}.  

Importantly, experiments have shown that a plasma with NT shaping exhibits improved confinement over PT L-mode plasmas with otherwise similar parameters (current, magnetic field, density, and auxiliary heating) in both DIII-D \cite{austin2019achievement,Marinoni_2019,Marinoni_2021} and TCV \cite{Pochelon_1999, Coda_2022, Camenen_2007}. This improved confinement is attained without entering H-mode, so $\sub{P}{SOL}$ is not required to be greater than $\sub{P}{LH}$ as it would be in PT H-mode, which allows for the use of techniques like seeding noble gas impurities in NT to further lower $\sub{P}{SOL}$ while maintaining plasma performance \cite{ARCH_2022, CasaliSpecialIssue, EldonSpecialIssue}. Recent experiments on DIII-D have also extended the observed operating space in NT to reactor-relevant levels in non-dimensional parameters \cite{Thome2024, pazsoldan2023}. In a diverted configuration, DIII-D NT plasmas simultaneously exhibited $\sub{\beta}{N} > 3$, $\sub{f}{Gr} > 1$, and $\sub{q}{95} < 3$ with $\sub{H}{98y2} > 1$ where $\sub{\beta}{N}$ is the normalized plasma beta, $\sub{f}{Gr}$ is the Greenwald fraction \cite{Greenwald_density}, $\sub{q}{95}$ is the safety factor at $\sub{\psi}{N} = 0.95$, and $\sub{H}{98y2}$ is normalized confinement time from the ITBP H98y2 energy confinement scaling law \cite{ITER_98y2}. Additionally, recent gyrokinetic simulation work has found the reduction in turbulent transport in NT to be independent of machine size \cite{DiGiannatale_2024}, further encouraging the use of NT in a reactor-class device.

While a few NT FPP designs have been proposed at varying levels of fidelity \cite{MANTA, Medvedev_2015, Kikuchi_2019}, these studies primarily focused on one design point and the performance trade-offs between various input parameters have not yet been fully established. To provide greater context for these trade-offs, we evaluate the performance of a reactor-class NT tokamak around two published operating points: the MANTA design \cite{MANTA} and the larger design from Medvedev \textit{et al} \cite{Medvedev_2015}. We accomplish this by using the STEP code \cite{MeneghiniSTEP}, which facilitates predictive integrated modeling through easy and self-consistent data transfer between various equilibrium, stability, and transport codes. The two NT design points studied in this work differ most notably in size, magnetic field, and current, as described in section~\ref{sec:methods}. Specifically, we investigate changes to $\sub{P}{fus}$ and $\sub{P}{SOL}$ that result from changes in temperature and density profiles, auxiliary power $\sub{P}{aux}$, seeded impurity fraction $\sub{f}{imp}$, triangularity $\delta$, and major radius $\sub{R}{maj}$. The core effects from changes in toroidal magnetic field $\sub{B}{t}$, volume, and plasma current $\sub{I}{P}$ will be investigated through comparison of the smaller volume, higher magnetic field, lower current MANTA design and the larger volume, lower magnetic field, higher current Medvedev design. Of particular interest to us is the effect of the edge pressure boundary condition on fusion performance in NT, as a full characterization of the NT edge is currently absent from the literature. 

In section \ref{sec:methods}, we introduce the high-field (MANTA-like \cite{MANTA}) and high-volume (Medvedev-like \cite{Medvedev_2015}) base cases around which we analyze NT reactor performance. We describe the integrated modeling workflow used with the STEP code \cite{MeneghiniSTEP} for the majority of simulations in this work. The density profile changes from a reactor-relevant particle source, i.e., a particle source localized toward the plasma edge, are investigated. We determine that the Angioni scaling \cite{Angioniscaling} predicts a density peaking similar to that which would be evolved to in TGYRO from a near-edge particle source. Thus we implement a psuedo-evolving scheme for density for simplicity in subsequent scans, as described in section \ref{subsec:density_evolution}. In section $\ref{sec:geometry}$, we discuss $\sub{P}{fus}$ density changes from $\delta$ and $\sub{R}{maj}$ scans in a high-field core, and find that geometry is not as leveraging as the electron pressure at $\rho = 0.8$ $\sub{p}{e,0.8}$. In section \ref{sec:edge_characterization}, we discuss the NT edge and its present uncertainty. Due to the lack of a physics-based predictive model of an NT edge, we scan various temperature and density edge boundary conditions ($\sub{T}{e,0.8}$ and $\sub{n}{e,0.8}$, respectively) and find that $\sub{P}{fus}$ and $\sub{P}{SOL}$ are both highly dependent on both $\sub{T}{e,0.8}$ and $\sub{n}{e,0.8}$. We use infinite-$n$ ballooning stability codes on the region beyond TGYRO evolution ($\rho = 0.8$ to $\rho = 1.0$) to determine that a $\sub{H}{98y2} \approx 1$ (confinement is what would be expected from a PT H-mode plasma with similar parameters) and $\sub{f}{Gr} \approx 1$ high-field NT operating point is likely feasible from a ballooning stability standpoint for a variety of potential pedestal widths. In section \ref{sec:heating}, we evaluate the impact of $\sub{P}{aux}$ on $\sub{T}{e,0.8}$, and find that a relatively high temperature boundary condition is required for sufficient $\sub{P}{fus}$ and cannot be overcome by additional $\sub{P}{aux}$. We compare the performance of NT reactor-like core scenarios at various $\sub{T},{e,0.8}$ and $\sub{P}{aux}$ to other published FPP concepts. In section \ref{sec:impurity}, we determine that at levels of intrinsic impurities (helium, tungsten) assumed by other FPP designs, there is minimal effect on fusion power when compared to the effect of seeded impurities utilized in this work (krypton). Including krypton at various concentrations results in a nearly linear downward trend in $\sub{P}{SOL}$ from increased impurity fraction and a potential for using impurity fraction to optimize $\sub{P}{fus}$.

\begin{table*}[t]
\begin{center}
\caption{Summary of select approaches to a tokamak FPP.}
\label{tab:FPPparams}
\begin{tabular}{lcccc}
       & MANTA \cite{MANTA} & ARC \cite{ARC_2015,KUANG2018} & ARIES-ACT2 \cite{ARIES_ACT2} & EU-DEMO \cite{FEDERICI_DEMO} \\
    \hline $\sub{B}{t}$ (T) & 11     & 9.2   & 8.75 & 5.2  \\
            $\sub{I}{P}$ (MA)& 10    & 7.8   & 14   & 20\\
            $\sub{R}{maj}$ (m) & 4.6 & 3.3   & 9.75 & 9  \\
            $\sub{a}{minor}$ (m) & 1.2 & 1.1 & 2.44 & 3   \\
            $\sub{\delta}{sep}$  & -0.5  & 0.375 &  0.63    & \\
            $\sub{\beta}{N}$ &  1.25 & 2.6   & 2.6  &  2.6 \\
            $\sub{H}{98y2}$  &  0.79 & 1.8   & 1.22 & 1.5    \\
            $\sub{P}{fus}$ (MW)  &  451  & 525   & 2600 & 1800    \\
            $\sub{P}{H\&CD}$ (MW) &   39  & 39    & 106  & 50    \\
            $\sub{P}{SOL}$ (MW)  &   24  & $\sim94$ & $\sim336$ & $\sim260$
\end{tabular}
\end{center}
\end{table*}

\section{Simulation setup and methods}
\label{sec:methods}

\subsection{Establishing a high-volume and a high-field NT reactor base case}
\label{subsec:basecase}

There are currently two main strategies to reach reactor-relevant performance in tokamaks: the high-volume approach and the high-field approach.
For PT H-mode, two notable FPP designs utilizing the high-volume approach are ARIES-ACT2 \cite{ARIES_ACT2} and EU-DEMO \cite{FEDERICI_DEMO}. Meanwhile, the high-field approach is well-represented by ARC-class devices \cite{ARC_2015} and generally by the SPARC project \cite{Creely2023}. For reference, basic parameters describing these concepts are displayed in table \ref{tab:FPPparams}. In this work, we initialize a base case for both strategies applied to NT, using work by Medvedev \textit{et al} \cite{Medvedev_2015} and the MANTA collaboration \textit{et al} \cite{MANTA} as starting points for the high-volume and high-field strategies, respectively. This enables us to evaluate advantages and disadvantages of both strategies in the NT operating space.

MANTA (Modular, Adjustable, Negative Triangularity ARC) is a high-field ($\sub{B}{t} \approx 11 \mathrm{T}$) NT FPP design. It is compliant with requirements laid out in the National Academy of Science, Engineering and Medicine’s (NASEM) report ``Bringing Fusion to the U.S. Grid” \cite{NASEM_report}, made possible in part by utilizing a FLiBe liquid immersion blanket, demountable HTS magnet joints, seeded krypton, and an NT core that exhibits sufficiently high confinement without ELMs. The simple divertor design of MANTA requires that $\sub{P}{SOL}$ be below 40 MW for a separatrix density of $0.9 \times 10^{20}/\mathrm{m}^3$ \cite{Miller2024}. Fusion power is additionally constrained to be within 400-500 MW when using 40 MW of auxiliary heating to meet the NASEM net electric goal of $\geq 50$ MWe \cite{MANTA}. 

Compared to MANTA, the design outlined in Medvedev \textit{et al.} is a lower field ($B_\mathrm{t} \approx 6 \mathrm{T}$) larger major radius ($R_\mathrm{maj} \approx 7 \mathrm{m}$) NT tokamak \cite{Medvedev_2015}. The primary focus of \cite{Medvedev_2015} was the MHD stability of a theoretical NT reactor-class design. As such, only the pressure profile is reported in \cite{Medvedev_2015}; the density and temperature profiles are not explicitly constrained. To facilitate a direct comparison of a high-volume NT case with a high-field NT case, we use the PRO-create module in OMFIT \cite{MeneghiniOMFIT} to initialize similar profiles in the high-volume case to those in MANTA. Due to the high volume, the separatrix density had to be lowered significantly to avoid exceeding the Greenwald limit \cite{Greenwald_density}, and consequently the $\sub{n}{e,0.8}$ scans performed in section \ref{sec:edge_characterization} are over lower values than in the high-field case.

\begin{figure}[h]
    \centering
    \includegraphics[width=0.5\textwidth]{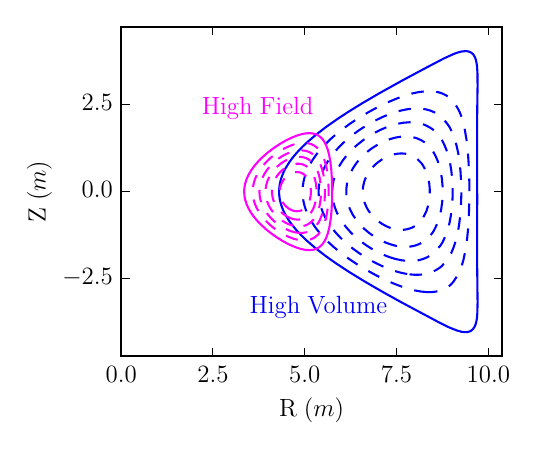}
    \caption{The equilibrium cross sections of the NT FPP reference cases considered in this work are plotted in $R$ and $Z$ coordinates. The high-field case ($\sub{B}{t} = 11$ T) is in pink and the high-volume case ($\sub{B}{t} = 6.2$ T) is in blue.}
    \label{fig:MANTA_vs_Medvedev}
\end{figure}

All scans in this work are done around either the high-volume or a high-field operating points with $\sub{H}{98y2} \approx 1$ and $\sub{f}{Gr} \approx 1$ outlined in table \ref{tab:MANTA_vs_Medvedev}. Parameters taken directly from references \cite{MANTA} and \cite{Medvedev_2015} are in bold. Figure \ref{fig:MANTA_vs_Medvedev} shows the equilibrium cross sections of both reference cases in $R$ and $Z$ coordinates. As will be discussed in section \ref{sec:edge_characterization}, there is significant uncertainty in predicting the edge condition for an NT FPP. To partially accommodate this uncertainty, we restrict the edge of our high-volume and high-field base cases via upper bounds on the normalized parameters $\sub{H}{98y2}$ and $\sub{f}{Gr}$ because they are both largely affected by the edge pressure condition in any plasma. Reactor-class fusion devices may be able to operate at $\sub{f}{Gr}>1$ because they will exhibit higher power densities than current devices \cite{Buttery_2021}. Supporting this idea, non-ELMing NT plasmas have accessed $\sub{H}{98y2} > 1$ simultaneously with $\sub{f}{Gr} > 1$ on DIII-D \cite{pazsoldan2023, Thome2024}. For these reasons, we chose to establish the high-volume and high-field base cases with normalized confinement time and density $H_\mathrm{98y2} \approx 1$ and $\sub{f}{Gr} \approx 1$, respectively. Obtaining $\sub{H}{98y2} \approx 1$ and $\sub{f}{Gr} \approx 1$ is accomplished by varying the temperature and density at $\rho = 0.8$ until a converged solution is found, where the method to find a converged solution is described below in subsection \ref{subsec:STEP}. 

\begin{table}[h!]
\begin{center}
\caption{Base-case parameters for the high-field (MANTA-like) and high-volume (Medvedev-like) operating points. Bolded numbers are the parameters taken directly from the original MANTA \cite{MANTA} and Medvedev \cite{Medvedev_2015} configurations. Brackets indicate volume average.}
\label{tab:MANTA_vs_Medvedev}
\begin{tabular}{c|c|c}
                            & High-field & High-volume \\
    \hline\hline $\sub{B}{t}$ (T) & \textbf{11}    & \textbf{6.2}      \\
    \hline $\sub{I}{P}$ (MA)& \textbf{10}    & \textbf{15}       \\
    \hline $\sub{R}{maj}$ (m) & \textbf{4.6} & \textbf{7.0}     \\
    \hline $\sub{a}{minor}$ (m) & \textbf{1.2} & \textbf{2.7}    \\
    \hline $\delta$         & \textbf{-0.5}  & \textbf{-0.9}     \\
    \hline $\kappa$         & \textbf{1.4}   & \textbf{1.5}      \\
    \hline $\sub{\beta}{N}$ &       1.9      &    1.7          \\
    \hline $\sub{q}{95}$    &     2.6        &    3.5          \\
    \hline $\sub{P}{aux}$ (MW)  & \textbf{40}    & 40       \\
    \hline $\sub{P}{fus}$ (MW)  & 966   &  879       \\
    \hline $\sub{P}{SOL}$ (MW) & 91 & 149            \\
    \hline $\sub{P}{rad}$ (MW) & 142   & 67             \\
    \hline $\sub{H}{98y2}$ & 1.0  &  0.95                 \\
    \hline $\sub{f}{Gr}$ & 0.94 &  1.05                   \\
    \hline $\langle{n}\rangle$ ($10^{20}/m^3$) & 4.1  &  1.4     \\
    \hline $\langle\sub{T}{e}\rangle$ (keV) &  9.4 &     9.2      \\
    \hline $\langle\sub{T}{i}\rangle$ (keV) & 9.1  &       8.8    \\
    \hline $\sub{n}{e,0.8}$ ($10^{20}/m^3$) & 1.8 & 0.58 \\
    \hline $\sub{T}{e,0.8}$ (keV) & 6.8 &  6.4       \\

\end{tabular}
\end{center}
\end{table}

\subsection{Integrated modeling with the STEP code}
\label{subsec:STEP}

\begin{figure}[h]
    \centering
    \includegraphics[width=0.5\textwidth]{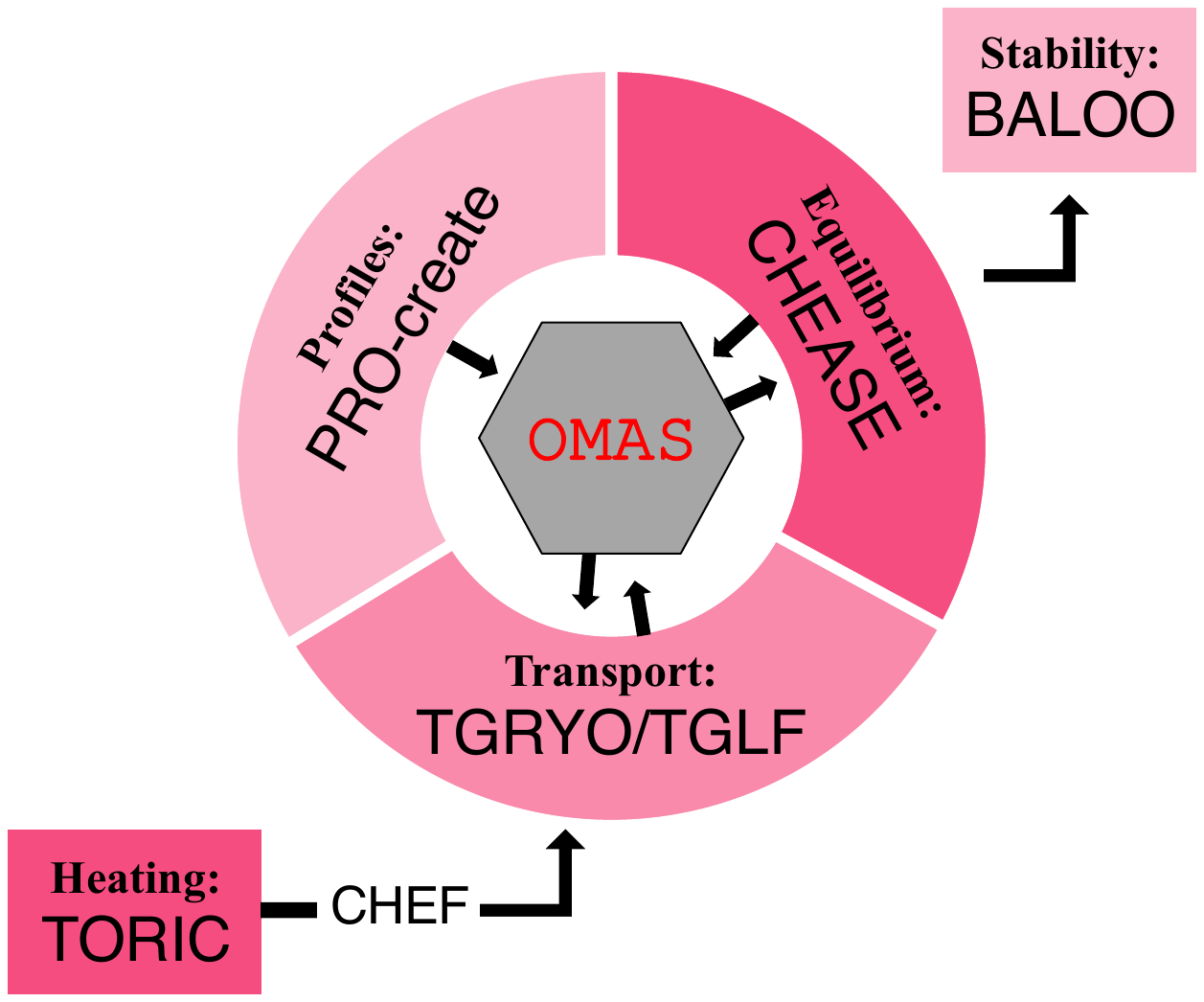}
    \caption{A diagram of the workflow used in this work. Profiles are created in PRO-create. The outputs of codes CHEASE and TGYRO are converted to the OMAS data structure and to pass between codes for self-consistency. Heat deposition profiles from TORIC are copied into CHEF for easy integration into the STEP workflow. Equilibria from CHEASE are passed into BALOO for ballooning stability analysis.}
    \label{fig:STEP_workflow}
\end{figure}

The integrated modeling in this work was performed with the STEP (Stability, Transport, Edge, Pedestal) code \cite{MeneghiniSTEP}. STEP enables self-consistent iteration between various OMFIT \cite{MeneghiniOMFIT} modules. We utilize CHEASE \cite{LutjensCHEASE} for equilibrium calculations, TGYRO \cite{CandyTGYRO} with TGLF \cite{StaeblerTGLF} for transport, and BALOO \cite{MillerBALOO} for infinite-$n$ ballooning instability. It is of note that bootstrap current was not included in the simulations in this work except in the discussion of ballooning stability in the edge in subsection \ref{subsec:ballooning}, but the inclusion of bootstrap current is not expected to alter the results of the core significantly. TORIC \cite{TORIC} was used on the MANTA core scenario to provide heat deposition profiles which were passed to the STEP workflow through the CHEF \cite{LyonsCHEF} module. A diagram of this workflow is shown in figure~\ref{fig:STEP_workflow}. TORIC was not used to solve for heating in the high-volume case. Heating for the high-volume case was instead copied from the heat deposition profiles solved for in MANTA and scaled as needed. TGYRO is a physics-based transport solver that uses NEO for neoclassical transport calculations and TGLF for turbulent transport calculations with moderate computational cost.

The modules mentioned thus far are primarily made for and/or trained on PT plasmas, and as such may not capture all NT effects. While the geometry change is taken into account in terms of surface area and volume, we note that TGLF is not a full gyrokinetic model, so there may be gyrokinetic effects of NT shaping that are not accounted for. However, gyrokinetic analysis done in other work has implied that the improved confinement of NT is likely due to gradients beyond $\rho = 0.8$ \cite{Aucone_2024,Mariani_2024}, where TGYRO evolution does not extend to in this work. Gradients in this region are subject to stronger triangularity shaping, as toroidal plasmas become more circular farther from the edge. For example, in the high-field base case, triangularity increases from about -0.5 to -0.3 from $\rho = 1.0$ to $\rho = 0.8$. Even so, reference \cite{DiGiannatale_2024} found that even though the effect is stronger at high radii, NT exhibits reduced transport over PT at low radii as well. Additionally, TGYRO/TGLF has been shown to reliably recreate DIII-D NT shots with various saturation rules \cite{Sciortino_2022,McClenaghan2024}, increasing the confidence with which we can apply these models to a reactor concept. However, we primarily use the SAT-2 saturation rule because it has been shown to better match experiment than SAT-0 at high powers and includes geometry effects \cite{Staebler_2021}. Unless otherwise stated, all simulations in this work evolve the temperature profiles from $\rho = 0.8$ to $\rho = 0$ with TGYRO/TGLF, using the ICRH heating profiles that were simulated for MANTA in TORIC with CQL3D\cite{cql3d} (with scaling as needed) and holding all density profiles constant with electron density profiles at the Angioni peaking $\sub{pk}{Angioni}$ \cite{Angioniscaling} and impurity density profiles scaled with electron density. Krypton is also included at a fraction of 0.001 as a seeded impurity with an otherwise 50/50 D/T fuel mix in all simulations unless otherwise noted. The use of impurities is elaborated on in section \ref{sec:impurity}. 

Only simulations fully converged in TGYRO are shown in this work. The definition of ``full" convergence for purposes of this work is described in \ref{apx:convergence}.  

\subsection{Density psuedo-evolution using density peaking scaling law}
\label{subsec:density_evolution}
Most currently operating tokamaks are fueled using neutral beam injection (NBI). NBI allows on-axis particle sourcing, which can increase density peaking and overall aides the tokamak's performance \cite{StorkNBI}. However, due to the relatively large size and high-field of a reactor-class tokamak, fueling by NBI may not be practical in an FPP \cite{Hopf_2021,HemsworthNBI}. Instead, the particle sources in a reactor are likely to be confined to the edge region, outside of $\rho = 0.8$. 

\begin{figure}[h]
    \centering
    \includegraphics[width=0.5\textwidth]{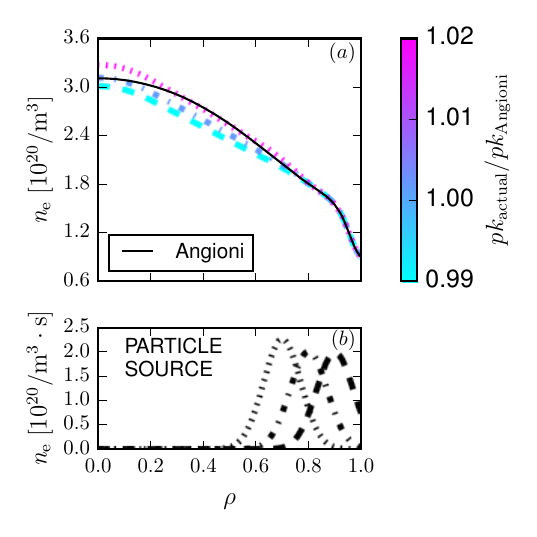}
    \caption{The variation in converged electron density profiles in the high-field case as a result of scanning a Gaussian particle source from $\rho = 0.7$ to $\rho = 0.9$. Electron sources are shown on bottom, and all integrate to the same number of electrons. The dashed lines correspond between the top and bottom plots. The colorbar gives the ratio of the actual density peaking $\sub{pk}{actual}$ to the Angioni predicted peaking $\sub{pk}{Angioni}$.}
    \label{fig:particlesourcechange}
\end{figure}

To assess the effect on density peaking $pk$ of a particle source in this region, we scan a Gaussian electron source from $\rho = 0.7$ to $\rho = 0.9$ in a MANTA-like core scenario. In these scans, TGYRO/TGLF with SAT-0 converges the density profile to one that has no more than a 2\% deviation from the peaking predicted by the Angioni scaling \cite{Angioniscaling} $\sub{pk}{Angioni}$, as shown in figure \ref{fig:particlesourcechange}. SAT-0 was used in this case because it is known to be easier to converge particle flux and heat flux simultaneously. Given the results of our particle source scans we suspect that TGYRO/TGLF would predict a similar peaking as is expected from the Angioni scaling. Though it is a challenge to converge particle flux and heat flux simultaneously with SAT-2, we chose to use SAT-2 in the remainder of this work because it includes additional geometry effects that SAT-0 and SAT-1 do not. We therefore use the $\sub{pk}{Angioni}$ prediction for density and TGYRO/TGLF with SAT-2 for temperature profile prediction in all subsequent parameter scans for ease. This density peaking prediction is dependent on collisionality, NBI source, and normalized plasma pressure $\beta$. As both the collisionality and $\beta$ are dependent on temperature, the peaking prediction changes as the temperature profile evolves in TGYRO. As a result, we pseudo-evolve the density between iterations by adjusting the core density after running TGYRO/TGLF and repeating until we find a converged solution with density peaking within the Angioni prediction. The reference ``Angioni" profile in figure \ref{fig:particlesourcechange} is that which was predicted using this procedure. 

It is of note that the Angioni scaling only uses a dataset of AUG and JET H-mode observations. However, it was found that collisionality was the most statistically significant parameter in the analysis \cite{Angioniscaling}. The normalized collisionality of MANTA is 0.4 at $\rho = 0.9$ and drops monotonically to 0.02 in the core. Note that the collisionality is high in the edge, which is not allowed in PT H-mode reactor core scenarios. However, NT does not have the same current-limiting pedestal demands as PT H-mode \cite{Paz-Soldan_2021, MANTA, Marinoni_2021}.

\section{Impact of the NT pressure boundary condition variation on fusion performance}
\label{sec:edge_characterization}
A significant remaining uncertainty in NT performance prediction is establishing the proper edge pressure condition. This is because the NT edge region is unique in that it is not a true ``L-mode" or ``H-mode" edge. Experiments have shown that NT can have steeper gradients than PT L-mode in the region outside of $\psi_\mathrm{N} = 0.9$, forming a small ``pedestal" while remaining ELM-free \cite{Marinoni_2021, Nelson_2022, Nelson_2023, Nelson2024}. However, they are still often able to recover the same pressure as PT H-mode in the core \cite{austin2019achievement, Coda_2022, pazsoldan2023, Marinoni_2021}. Because the NT edge has yet to be characterized to the extent of the PT H-mode edge, which still itself retains significant uncertainty during extrapolations to an FPP, there is more uncertainty around what pressures and pressure gradients could potentially be obtained in an NT reactor edge. 

\begin{figure}
    \centering
    \includegraphics[width=0.5\textwidth]{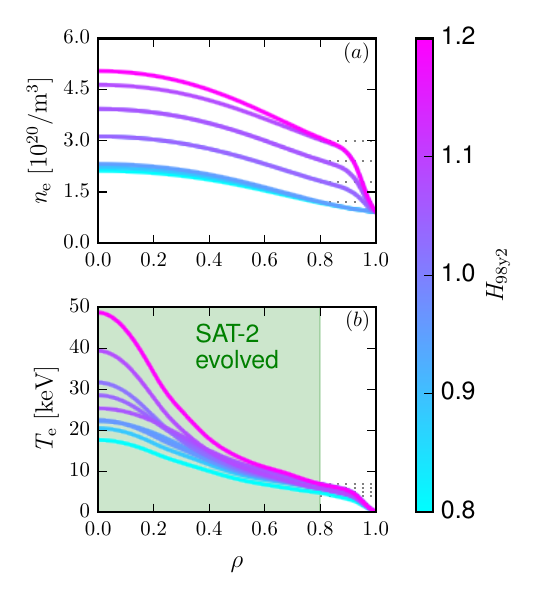}
    \labelphantom{fig:Angioni_evolving-a}
    \labelphantom{fig:Angioni_evolving-b}
    \caption{Electron density and TGYRO converged temperature profiles for boundary condition scans in the high-field case. Density profiles are at the Angioni peaking and temperature profiles are evolved in TGYRO with SAT-2 from $\rho = 0$ to $\rho = 0.8$. Dotted gray lines are at the values scanned at $\rho = 0.8$. Colorbar gives $H_\mathrm{98y2}$.}
    \label{fig:Angioni_evolving}
\end{figure}

\begin{figure}
    \centering
    \includegraphics[width=0.5\textwidth]{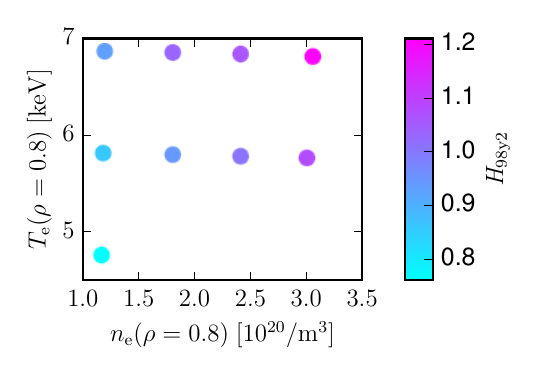}
    \caption{A representation of edge scans performed over $\sub{T}{e,0.8}$ and $\sub{n}{e,0.8}$. Each point represents a converged simulation as described in section \ref{sec:methods}. Colorbar gives $\sub{H}{98y2}$.}
    \label{fig:MANTA_edge_scan_dots}
\end{figure}

To capture effects related to the edge boundary condition, we employ a ``brute force" characterization of the edge in both the high-field and high-volume configurations. This characterization is performed by scanning over four electron temperature and four electron density boundary conditions at $\rho = 0.8$, for a total of 16 simulations. We use the psuedo-density evolving method described in section \ref{sec:methods} and evolve temperature from $\rho = 0$ to $\rho = 0.8$ with TGYRO/TGLF and SAT-2. The resultant electron density and converged temperature profiles are shown in figure \ref{fig:Angioni_evolving} with boundary condition values shown by the dotted gray lines. These values can also be seen from the x and y axis in figure \ref{fig:MANTA_edge_scan_dots}. In figure \ref{fig:Angioni_evolving} and \ref{fig:MANTA_edge_scan_dots}, $\sub{H}{98y2}$ is given by the colorbar. The green region in figure \ref{fig:Angioni_evolving-b} is that which was evolved in TGYRO/TGLF. Note that as the edge value increases in both electron temperature and density, $\sub{H}{98y2}$ increases as well. The highest pressure boundary condition for the high-field configuration tested in this work exhibited a confinement time $20\%$ over that expected from the $\sub{\tau}{98y2}$ scaling \cite{ITER_98y2}, while the highest for the high-volume configuration was $5\%$ below. Attaining $\sub{H}{98y2} > 1$ for both cases is difficult at low pressure boundary conditions.

\begin{figure}
    \centering
    \includegraphics[width=0.5\textwidth]{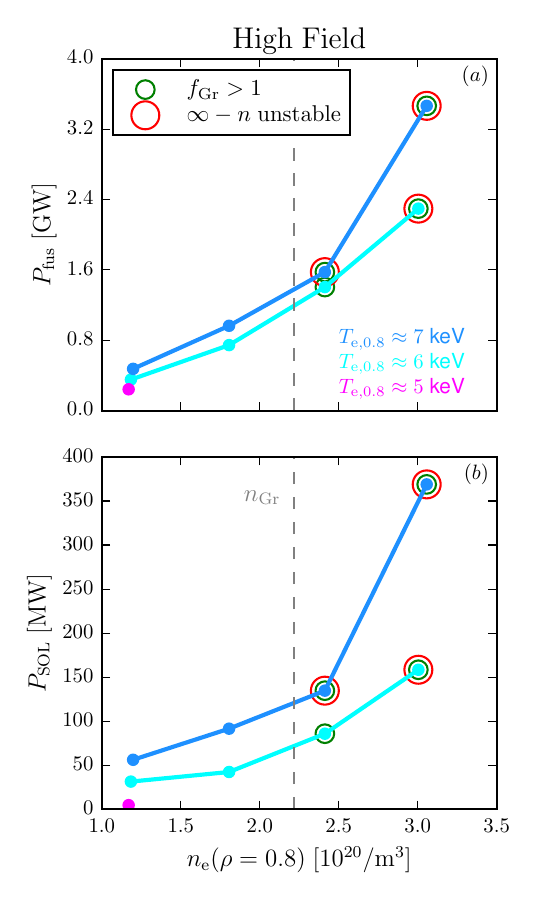}
    \caption{$\sub{P}{fus}$ and $\sub{P}{SOL}$ increase with electron density and temperature boundary conditions for the high-field case.  Each point represents a converged simulation as described in section \ref{sec:methods}. Simulations where the Greenwald fraction \cite{Greenwald_density} exceeded 1 (calculated using volume average) are circled in green. Simulations that are infinite-$n$ ballooning unstable in the edge with a pedestal width of 0.1 are circled in red. The Greenwald number is shown by the dotted gray line.}
    \label{fig:MANTA_edge_scan}
\end{figure}

\begin{figure}
    \centering
    \includegraphics[width=0.5\textwidth]{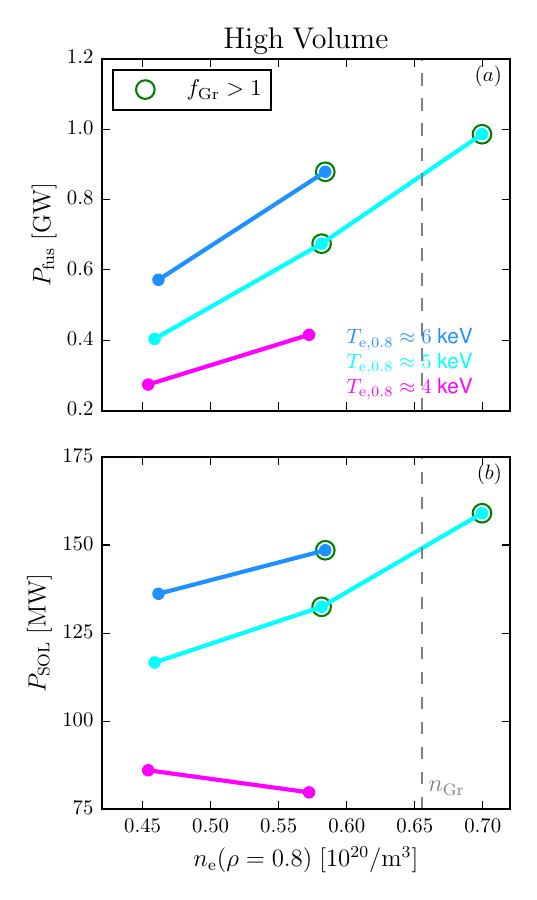}
    \caption{$\sub{P}{fus}$ increases with electron temperature and density boundary conditions for the high-volume case. $\sub{P}{SOL}$ increases with increasing electron temperature boundary condition but decreases with electron density boundary condition for the $\sub{T}{e,0.8} \approx 4$ keV points. Each point represents a converged simulation as described in section \ref{sec:methods}. Simulations where the Greenwald fraction \cite{Greenwald_density} exceeded 1 (calculated using volume average) are circled in green. The Greenwald number is shown by the dotted gray line. None of these cases were infinite-$n$ ballooning unstable with a pedestal width of 0.1.}
    \label{fig:Medvedev_pedscan}
\end{figure}

In both the high-field and high-volume cases, fusion performance is seen to depend heavily on the pressure boundary condition at $\rho = 0.8$. This can be seen in figures \ref{fig:MANTA_edge_scan} and \ref{fig:Medvedev_pedscan} for the high-field and high-volume cases, respectively. The same conclusion for high-field was drawn in reference \cite{SPARC_Lmode} from high fidelity modeling of PT L-mode operation in SPARC. Investigation of the edge physics in tokamak plasmas requires further work in both cases.

In both figure \ref{fig:MANTA_edge_scan} and figure \ref{fig:Medvedev_pedscan}, the top plot shows $\sub{P}{fus}$ increasing with increased electron density at $\rho = 0.8$ ($\sub{n}{e,0.8}$) for three distinct temperature values at $\rho = 0.8$ ($\sub{T}{e,0.8}$). Note that $\sub{P}{fus}$ also increases with increased $\sub{T}{e,0.8}$. The bottom plot in both figures shows the same trend for $\sub{P}{SOL}$ for both cases except for at the lowest temperature in the high-volume case (blue line in figure~ \ref{fig:Medvedev_pedscan}). All points shown are converged using the psuedo-evolving scheme for density and TYGRO/TGLF for temperature as described in section \ref{sec:methods}. The green circled points in figure \ref{fig:MANTA_edge_scan} and figure \ref{fig:Medvedev_pedscan} indicate simulations in which the Greenwald fraction (calculated with volume-averaged density) exceeded unity. The red circled points in figure \ref{fig:MANTA_edge_scan} are those in which the infinite-$n$ ballooning stability limit was exceeded in the region from $\rho = 0.8$ to $\rho = 1.0$ for a pedestal width of 0.1 in $\sub{\psi}{N}$. The scans over $\sub{n}{e,0.8}$ are at much smaller values for the high-volume case than for the high-field case due to exceeding the Greenwald limit at higher densities. This results in higher $\sub{P}{SOL}$ for the high-volume case as well, due to the lower separatrix density required. Therefore, a significant benefit of the high-field case from a power handling standpoint is the ability to employ higher separatrix density over higher volume cases without exceeding the Greenwald limit. This is even with the high-volume case employing higher $\sub{I}{P}$, because the minor radius is more than twice that of the high-field case, which reduces the Greenwald limit significantly. Note also that the high-field case displays higher $\sub{P}{fus}$, but also converged with higher $\sub{T}{e,0.8}$ than the high-volume case. This is once again likely attributable to the higher $\sub{n}{e,0.8}$ employed in the high-field case. The mutual dependence of edge temperature and density and their significant effect on $\sub{P}{fus}$ and $\sub{P}{SOL}$ motivates additional work in characterizing the NT edge boundary condition in future experiments and modeling.

\subsection{Ballooning stability in the region outside of $\rho = 0.8$}
\label{subsec:ballooning}

In addition to the impact of pressure boundary condition on core fusion performance, another uncertainty related the edge in an NT FPP is the specific conditions under which an NT plasma remains ELM-free. The leading experimental explanation of ELM-free performance at significant negative triangularity is the closure of the second infinite-$n$ ballooning stability region, which restricts pressure gradient growth \cite{Nelson_2022, Nelson_2023}. In these experiments, where pressure gradients are confined to the first stability region, no ELMs are observed. However, it is of note that DIII-D NT experiments have also suggested that there is a gradient limiting mechanism that precedes ballooning instability \cite{Nelson_2023, Nelson2024}. Thus ballooning instability is likely only sufficient as an upper bound on the pressure gradient; it is possible that the pressure gradient will be even lower, which would lower global performance. This further justifies the scans of the edge pressure boundary condition performed earlier in this section. In this work, we use ballooning stability as a proxy for access to the ELM-free state, assuming that if the normalized pressure gradient remains in the first stability region the plasma will not generate ELMs, as suggested by reference \cite{Nelson_2022}. %
Gradients at the core-edge boundary ($\rho = 0.8$) in this work were maintained well below the infinite-$n$ ballooning stability limit in all scans.   %

\begin{figure}
    \centering
    \includegraphics[width=0.5\textwidth]{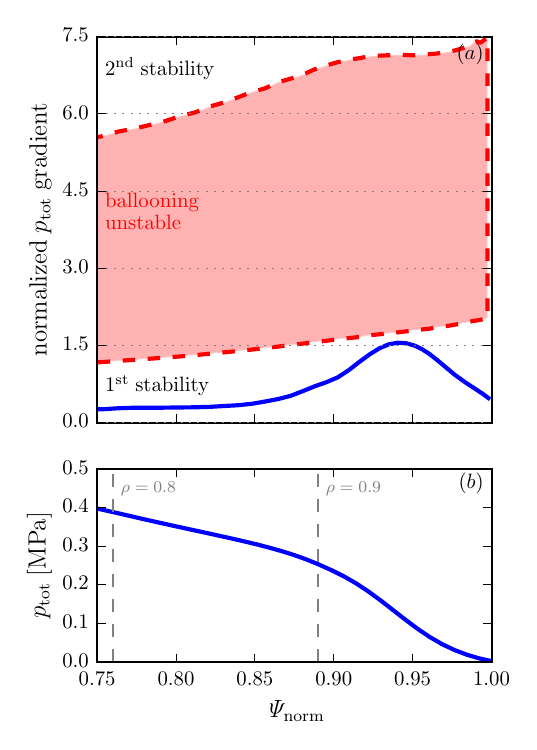}
    \labelphantom{fig:ballooning_stability-a}
    \labelphantom{fig:ballooning_stability-b}
    \caption{A ballooning stability diagram of the High $\sub{B}{t}$ case with $\sub{f}{Gr} \approx 1$ and $\sub{H}{98y2} \approx 1$. The normalized pressure gradient is in blue. The ballooning instability region is in red. The modeled pedestal for the high-volume case (not shown) is significantly below the first stability limit for ideal ballooning modes.}
    \label{fig:ballooning_stability}
\end{figure}

An example of the infinite-$n$ ballooning stability calculated by BALOO is given in figure \ref{fig:ballooning_stability} for the high-field case with $\sub{f}{Gr} \approx 1$ and $\sub{H}{98y2} \approx 1$ and a pedestal width of $\sub{\Delta}{ped} = 0.1$. Here the red region indicates the ballooning unstable region while the normalized pressure gradient and the total pressure in the edge are shown by the blue profiles in figures \ref{fig:ballooning_stability-a} and \ref{fig:ballooning_stability-b}, respectively. Since edge profile prediction is outside the scope of this work, we extrapolate from the end of the TGYRO evolution region ($\rho=0.8$) to $\rho = 1$ with an H-mode-like \texttt{tanh} pedestal shape. In figure \ref{fig:ballooning_stability-a} the pressure gradient remains in the first stability region even at the steepest point. Similar analysis of the high-volume case reveals a significantly larger gap between the first stability limit and the normalized pressure gradient, as the edge is instead limited by the imposed $f_\mathrm{Gr}=1$ constraint. Should Greenwald fractions greater than unity be achievable in an NT FPP, it may be possible to raise the edge pressure further in machines with larger $\sub{R}{maj}$ before destabilizing ideal ballooning modes \cite{Nelson_2022}. 

For ease of comparison with calculations presented in \cite{MANTA}, the bootstrap current is omitted from figure~\ref{fig:ballooning_stability-a}. However, we note that inclusion of this effect can impact the local magnetic shear in regions of strong pressure gradients, potentially leading to a reduction of the maximum ballooning-stable pressure gradient and adding uncertainty dependent on the choice of bootstrap model. As such, results presented in this work should be treated as a theoretical upper limit on the edge pressure gradient rather than a precise prediction, as mentioned above. The reduction of the maximum achievable edge pressure gradient with the inclusion of bootstrap current is less pronounced at larger $\sub{R}{maj}$, again suggesting that an increase in aspect ratio could be used to recover edge performance that may otherwise be limited by ballooning stability \cite{Nelson_2022}; the uncertainty in ballooning stability due to bootstrap current is more important at low aspect ratio. 

\begin{figure}
    \centering
    \includegraphics[width=0.5\textwidth]{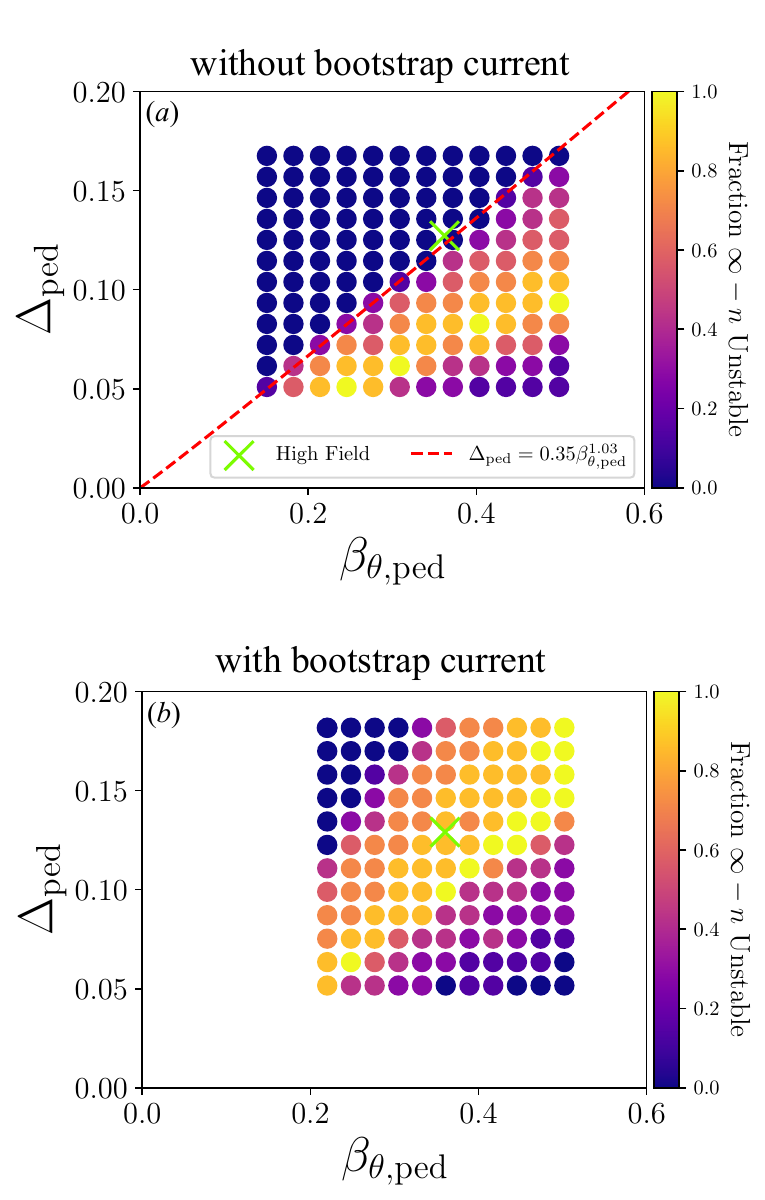}
    \labelphantom{fig:Parisi_stability-a}
    \labelphantom{fig:Parisi_stability-b}
    \caption{Pedestal width $\sub{\Delta}{ped}$ in normalized $\psi$ versus pedestal toroidal normalized pressure $\sub{\beta}{\theta,ped}$ for the high-field case. The colorbar shows the fraction of pedestal half-widths with a given $\sub{\Delta}{ped}$ and $\sub{\beta}{\theta,ped}$ that are unstable to infinite-$n$ ballooning modes. The red dashed line in (a) indicates the fit of the instability boundary for various pedestal size combinations. The green ``$\times$" indicates the high-field case equilibrium.}
    \label{fig:Parisi_stability}
\end{figure}

Beyond the conventional presentation of ballooning stability presented in figure~\ref{fig:ballooning_stability}, it can be informative to examine a scan of the pedestal width and height for each given case. This is especially true in NT scenarios where the pedestal is not well-modeled using H-mode pedestal predictions like EPED \cite{Nelson2024} and a physics-based predictive model capable of describing the pedestal width and height has yet to be developed. In figure \ref{fig:Parisi_stability}, we inspect the stability of potential NT edge pressure gradients resulting from various pedestal widths in the region from $\rho = 0.8$ to $\rho = 1.0$ by running gk\_ped \cite{parisi2024_gkped} on the high-field case. The gk\_ped code is a linear gyrokinetic threshold model that generates self-consistent equilibria (including the bootstrap current, calculated with the analytical formulae from \cite{redl_new_2021}) that vary in pedestal width, pedestal density, and pedestal temperature and then solves for the ballooning critical pedestal via the BALOO stability formalism at each point. Here we define the ballooning critical pedestal as the profile form that achieves the highest pressure gradient at a particular pedestal width before becoming ideal ballooning unstable. In figure~\ref{fig:Parisi_stability-a}, which shows the calculations without the inclusion of the bootstrap current, the ballooning critical pedestal is described by the relationship
\begin{equation}
    \sub{\Delta}{ped} = 0.35\sub{\beta}{\theta,ped}^{1.03},
    \label{eq:gkped}
\end{equation}
where $\sub{\Delta}{ped}$ is the pedestal width and $\sub{\beta}{\theta,ped}$ is the normalized poloidal pedestal pressure. The color bar in figure~\ref{fig:Parisi_stability} represents the fraction of radial locations of the pedestal half-width that are ballooning unstable. We note that, for the high-volume cases presented, the modeled scenarios presented in this work feature pedestals that lie below the ballooning critical pedestal calculated by gk\_ped.

Equation~\ref{eq:gkped} can be used to describe an upper ballooning-stable limit on the NT pedestal height as a function of pedestal width, as characterized by the traditional H-mode-like \texttt{tanh} pedestal shape. However, as seen in the comparison to figure~\ref{fig:Parisi_stability-b}, which includes an analytic model for the bootstrap current in the edge region, this ballooning critical pedestal may over-predict the edge gradient in NT FPPs. This discrepancy is potentially compounded by observations on DIII-D that the actual edge gradient in NT plasmas lies somewhere below the infinite-$n$ stability limit \cite{Nelson_2023, Nelson2024}, suggesting that an additional model is required for accurate prediction of the NT edge profile. However, we note that the location of high-field equilibrium point in the instability region in figure~\ref{fig:Parisi_stability-b} does not invalidate the value of pressure employed at $\rho = 0.8$ for the core profile modeling presented in the work. Indeed, it is possible to achieve a similar pedestal height as in figure~\ref{fig:Parisi_stability-a} while remaining stable to ballooning modes by going to a larger pedestal width. This highlights the need to develop a constraint for the NT pedestal width similar to the EPED model in H-mode scenarios, as the ballooning stability itself cannot fully constrain a pressure boundary condition at $\rho = 0.8$.

Because the region between $\rho = 0.8$ and $\rho = 1.0$ is beyond the TGYRO evolution in this work, any further analysis in this region is out of scope. However, figures \ref{fig:ballooning_stability} and \ref{fig:Parisi_stability} suggest that NT boundary condition at $\rho = 0.8$ may reach high pressures and pressure gradients, beyond those expected in a PT L-mode, without becoming ballooning unstable. This is supported by NT experiments on DIII-D that have observed increased edge pressure and pressure gradients in the NT edge without ELMs \cite{Nelson2024, Thome2024, Marinoni_2021, Coda_2022}, though we note that edge transport barrier mechanics and turbulence suppression remain an open area of research in NT. Figure \ref{fig:Parisi_stability} illustrates that even in a high-field case with $\sub{H}{98y2} \approx 1$ and $\sub{f}{Gr} \approx 1$, there are many ballooning-stable pedestal shapes that may be possible in NT. In particular, it is possible to achieve the same temperature and density boundary conditions at $\rho=0.8$ with reduced pedestal gradients by increasing the width of the NT pedestal, which is not constrained in this work. Thus the profile shown in \ref{fig:ballooning_stability-b} is just one of many possibilities of an NT pedestal shape given the pressure at $\rho = 0.8$ and we encourage further experimental endeavours to characterize this region. In particular, physics-based constraints on the pedestal width or on the functional form of the pedestal would be valuable for improved predictive capabilities for NT FPPs, as they would enable a significant reduction in the parameter space available for stability codes like gk\_ped.
\section{Relative effect of triangularity and major radius on fusion performance}
\label{sec:geometry}
The importance of the boundary pressure at $\rho = 0.8$ on plasma performance is also seen when exploring changes in fusion performance due to geometry. While scanning triangularity and major radius, we found that changes in fusion power density due to $\sub{R}{maj}$ did not dominate over changes due to boundary electron pressure $\sub{p}{e,0.8}$, as will be shown.

The motivation for studying a high-volume case is that $\sub{P}{fus}$ increases with increasing volume \cite{freidberg_plasma_2008}, so larger $R_\mathrm{maj}$ can be employed for lower field devices to achieve performance comparable with more compact high-field devices. In a high-field device like MANTA, compactness is prioritized because size is expected to be a major cost driver. However, there is an additional benefit of large $R_\mathrm{maj}$ in that it allows for a larger central solenoid, enabling increased flux swings and longer pulse lengths. Thus $R_\mathrm{maj}$ is an import parameter to optimize for FPP performance. In this work, we do not discuss the potential engineering benefits of increased $\sub{R}{maj}$ in an NT FPP, only the effect on transport and core performance.

Volume generally decreases with increasingly negative triangularity, which affects fusion power output. NT has shown robust avoidance of ELMs when $\delta < \delta_\mathrm{crit}$ in experiment, where $\delta_\mathrm{crit}$ is a critical triangularity that is device-dependent, around $\delta_\mathrm{crit}\sim-0.15$ on DIII-D \cite{Nelson_2022}. This $\delta_\mathrm{crit}$ is unknown experimentally in a MANTA-like device, so it is safer to assume stronger negative triangularity to ensure the plasma is robustly ELM-avoidant. However, plasmas with stronger NT are more vertically unstable, so $\delta$ is also a parameter ripe for optimization \cite{Guizzo2024, Nelson_vertControl}. 

\begin{figure}[H]
    \centering
    \includegraphics[width=0.5\textwidth]{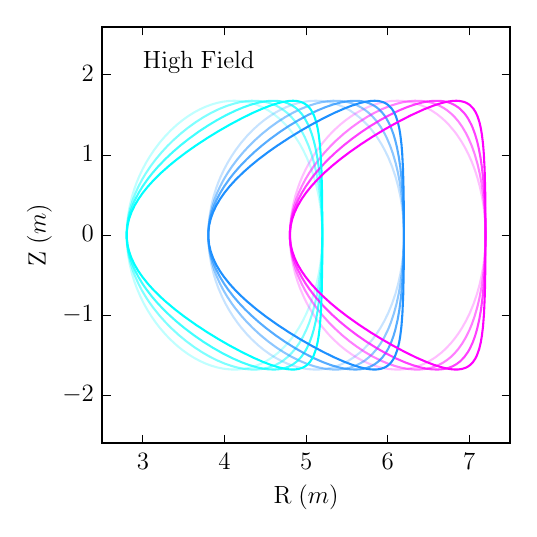}
    \caption{The last closed flux surfaces of equilibria scanned over various $\sub{R}{maj}$ and $\delta$ are plotted in $R$ and $Z$ coordinates. Each color represents a distinct $\sub{R}{maj}$ while increasing transparency of each contour indicates increasing $\delta$.}
    \label{fig:shapes}
\end{figure}

Interestingly, volume does not directly correlate with negative triangularity in these scans, but peaks at around $\delta = -0.3$. This is likely due to squareness not being held constant across all scans. We have plotted the last closed flux surface of the equilibrium shapes scanned in this section in figure \ref{fig:shapes} for reference. Each color is at a different major radius while the transparency of the contours increases with increasing $\delta$.

\begin{figure}[H]
    \centering
    \includegraphics[width=0.5\textwidth]{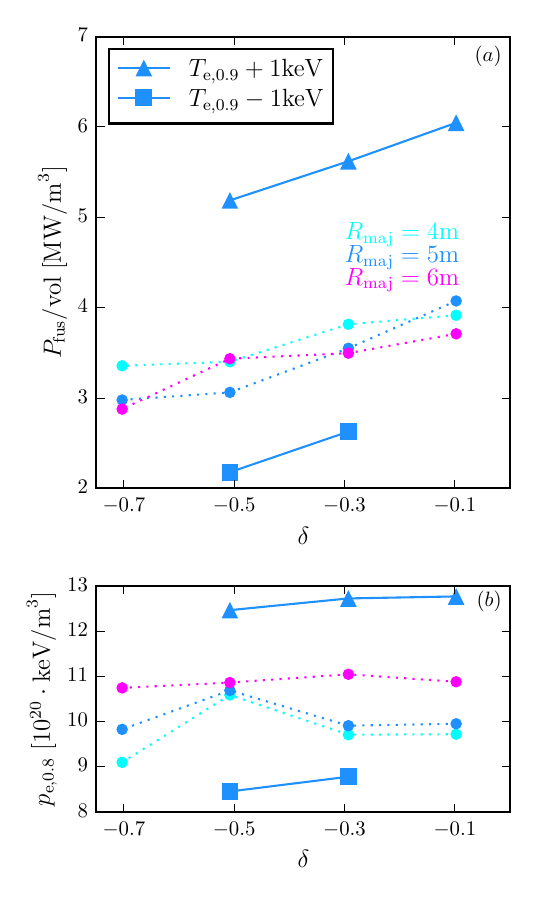}
    \labelphantom{fig:delta_scan-a}
    \labelphantom{fig:delta_scan-b}
    \caption{In (a), fusion power density is plotted versus triangularity for three distinct major radii. The dotted lines all target the same boundary temperature, and there is little variation in boundary pressure at $\rho = 0.8$ as shown in (b). The solid blue line with triangle markers targets a boundary temperature increased by 1 keV from that of the dotted line at $\sub{R}{maj} = 5$ m. The solid blue line with square markers targets a boundary temperature decreased by 1 keV from the dotted line at $\sub{R}{maj} = 5$  m. All points represent converged simulations.}
    \label{fig:delta_scan}
\end{figure}

To illustrate the impact of $\sub{R}{maj}$ and $\delta$ on fusion performance in an NT scenario, the fusion power density is plotted against $\delta$ in figure \ref{fig:delta_scan-a} with each distinctly colored line representing a different $\sub{R}{maj}$. We do not focus on the obvious increase in $\sub{P}{fus}$ with volume, instead investigating fusion power density to discover any underlying transport effects attributable to the change in shape. The dotted lines contain simulation points in which equilibria were initialized with the same $\sub{B}{t}$, $\sub{I}{P}$, $\sub{P}{aux}$, target $\sub{a}{minor}$, temperature profiles, and density profiles, with only $\delta$ and $\sub{R}{maj}$ changing between each simulation. Note that this does lead to changes in magnetic shear and safety factor $q$ as well, so there is potential performance optimization to be done at distinct $\delta$ and $\sub{R}{maj}$ owing to global stability considerations. For example, $q_\mathrm{95} = 2$ for the $\sub{R}{maj} = 6$ m cases here, which is the lower stability limit. Additional scans that could increase $\sub{I}{P}$ to maintain $q_\mathrm{95}$ constant at each $\sub{R}{maj}$ could expand the $\sub{R}{maj}$ optimization picture and are left to future work and should be combined with dedicated MHD stability modeling. Also note that scans of the shaping parameters lead to greater variation between the equilibrium in each simulation than in the $\sub{T}{e,0.8}$ and $\sub{n}{e,0.8}$ scans presented in section \ref{sec:edge_characterization}, leading to increased variation in equilibrium parameters that are targeted as inputs in CHEASE. In any case, an increase in the fusion power density with less negative triangularity is evident. This suggests that the optimal triangularity for operation of a NT FPP may be that which is just negative enough to avoid ELMs - further decrease in $\delta$ may result in a general decrease in the fusion power density.

In figure \ref{fig:delta_scan-b}, the electron pressure at $\rho = 0.8$ is plotted against $\delta$, highlighting some variation in edge pressure within each $\delta$ scan. In all TGYRO simulations in this work, the edge pressure at $\rho = 0.9$ was fixed while the pressure from $\rho = 0$ to $\rho = 0.8$ was allowed to evolve until convergence was met, resulting in some variation in edge pressure at $\rho = 0.8$. We tested the role of edge pressure on these results by performing two additional scans over $\delta$ at constant $\sub{R}{maj} = 5$ m. The first is shown by the solid blue triangle markers, with $\sub{T}{e,0.9}$ increased by 1 keV with all other parameters the same as the corresponding dashed $\sub{R}{maj} = 5$ m line in figure \ref{fig:delta_scan}(a). The second is shown by the solid blue square markers, with $\sub{T}{e,0.9}$ decreased by 1 keV. When decreasing $\sub{T}{e,0.9}$, it was more difficult to converge TGYRO in the edge for these cases. The $\sub{T}{e,0.9} + 1$ keV and $\sub{T}{e,0.9} - 1$ keV scans resulted in fusion power density changing more drastically than the change from $\sub{R}{maj} = 4$ m to $\sub{R}{maj} = 6$ m, but still displayed the same increase in fusion power density from increasing triangularity. Thus edge pressure has a significantly more profound effect on fusion power density than $\sub{R}{maj}$, though there is still the benefit of increased $\sub{P}{fus}$ from the increased volume at higher $\sub{R}{maj}$. 

Though there is increased vertical stability at weaker triangularity \cite{Guizzo2024} and increased $\sub{P}{fus}$ density at weaker triangularity, operating below a critical triangularity $\sub{\delta}{crit}$ will be necessary for ELM avoidance \cite{Nelson_2022, Nelson_2023}. For a MANTA-like device $\sub{\delta}{crit}$ cannot yet be experimentally verified, but it is clear that it would be beneficial to operate as close to $\sub{\delta}{crit}$ as possible while remaining ELM-free.

\section{The effect of the temperature edge condition on heating requirements}
\label{sec:heating}

\begin{figure}[h]
    \centering
    \includegraphics[width=0.5\textwidth]{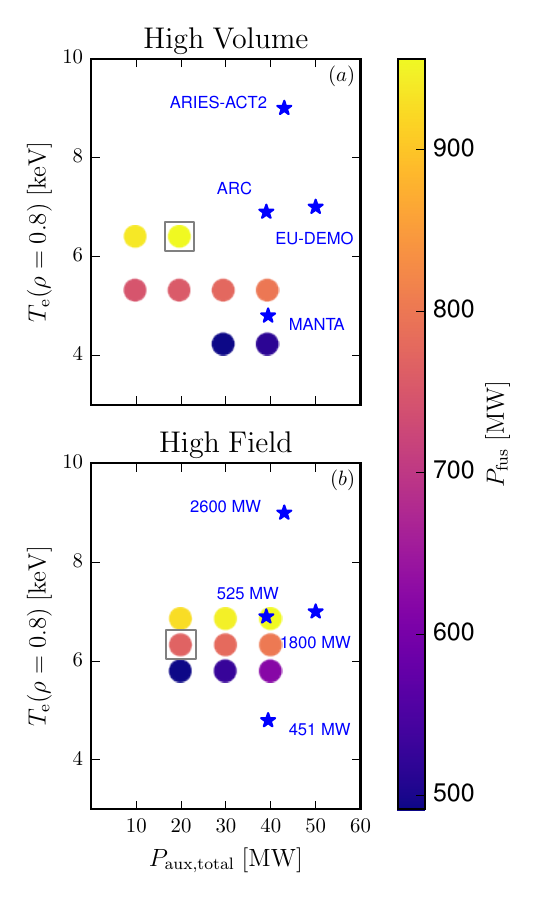}
    \caption{Electron temperature boundary condition at $\rho = 0.8$ versus total auxiliary power with colorbar representing fusion power in megawatts. Plot (a) is for the high-volume case and plot (b) is for the high-field case. Prevalent PT H-mode FPPS are represented by blue stars. The machine labels of the blue stars in (a) also correspond to the power labels of the blue stars in (b). The power labels of the blue stars in (b) give $\sub{P}{fus}$ of their respective devices.}
    \label{fig:Te_Paux_Pfus}
\end{figure}

Given the sensitivity of core performance on the temperature and density edge condition, establishing a set of reliable actuators and controls for these parameters is paramount to the successful design of an NT FPP. One potential path to control the temperature edge is to leverage auxiliary heating power $\sub{P}{aux}$. For MANTA, the full wave code TORIC was used with CQL3D\cite{cql3d} to determine 1D power deposition profiles for both ions and electrons from ICRH heating such that the total $\sub{P}{aux} \approx 40$ MW\cite{MANTA}. In $\sub{P}{aux}$ scans in this work, heating was not calculated self-consistently but instead was simply scaled from the MANTA power deposition profiles for both the high-field and high-volume cases. In figure \ref{fig:Te_Paux_Pfus}, scans of $\sub{T}{e,0.8}$ and $\sub{P}{aux}$ on the high-field and high-volume core are shown. The colorbar gives $\sub{P}{fus}$ in mega-watts, and selected tokamak FPP operating points are indicated by blue stars. Their corresponding parameters are given in table \ref{tab:FPPparams}.

It is clear in both cases that $\sub{P}{fus}$ increases with $\sub{T}{e,0.8}$. Though $\sub{P}{fus}$ also increases with $\sub{P}{aux}$, it is not as pronounced as the change due to $\sub{T}{e,0.8}$. Note that we cannot claim the level of $\sub{P}{aux}$ that will be required to maintain a certain $\sub{T}{e,0.8}$, as the relationship between $\sub{P}{aux}$ and $\sub{T}{e,0.8}$ is ultimately governed by edge physics. Instead, the two are varied independently in figure \ref{fig:Te_Paux_Pfus} to scope potential $\sub{P}{aux}$ and $\sub{T}{e,0.8}$ combinations and their corresponding $\sub{P}{fus}$. The relatively weak dependence of $\sub{P}{fus}$ on $\sub{P}{aux}$ suggests the possibility of higher gain solutions at a given $\sub{T}{e,0.8}$ by going to lower $\sub{P}{aux}$. We ultimately find that $\sub{T}{e,0.8}$ is more important than $\sub{P}{aux}$ in determining fusion performance, which once again motivates further investigation of the physics governing the NT edge.

Similar solutions can be found in both the high-field and high-volume cases at various $\sub{P}{aux}$, but are harder to converge at $\sub{P}{aux} < 10$ MW and $\sub{T}{e,0.8} < 5.5$ keV in the high-field case, requiring higher $\sub{T}{e,0.8}$ than the high-volume case to reach the same $\sub{P}{fus}$. Note that ARC and EU-DEMO, representing the high-field and high-volume path for PT H-mode FPPs, respectively, have approximately the same $\sub{T}{e,0.8}$ and $\sub{P}{aux}$ values. However, ARC has a fusion power of 525 MW while EU-DEMO has a fusion power of 1800 MW, due predominantly to its larger volume. If we compare the two points in grey squares on figure \ref{fig:Te_Paux_Pfus}, they are also at approximately the same $\sub{T}{e,0.8}$ and $\sub{P}{aux}$ value, but exhibit $\sub{P}{fus}$ of 786 MW for the high-field case and 940 MW for the high-volume case. While there are likely other parameters at play constituting the difference between ARC and EU-DEMO performance, NT appears to approximately follow a similar trend to PT H-mode between the high-field and high-volume approaches in this case. Note also that the zone in which the converged simulations lie also differs between the high-field and high-volume case in NT. This is due to the density difference between cases, with the high-volume case able to converge at a lower $\sub{n}{e,0.8}$. Referring to table \ref{tab:MANTA_vs_Medvedev}, $\sub{n}{e,0.8}$ for the high-field case is more than three times that of the high-volume case. Thus, a higher $\sub{T}{e,0.8}$ is required for power balance in TGYRO. The high-field case also displays higher sensitivity to $\sub{T}{e,0.8}$ than the high-volume case. 
The expected $\sub{T}{e,0.8}$ and $\sub{P}{aux}$ needed for comparable $\sub{P}{fus}$ to leading PT H-mode FPPs is comparable to that seen in these devices. ARC is the lowest shown here at $\sub{P}{fus} = 525$ MW and ARIES-ACT2 the highest at $\sub{P}{fus} = 2600$ MW, but note that ARIES-ACT2 has significantly higher $\sub{T}{e,0.8}$ which we've seen to be very influential on $\sub{P}{fus}$.

\section{Impurity analysis}
\label{sec:impurity}

Given the primary benefit of operating in NT is enhanced confinement without ELMs, we are interested in robustly \textit{avoiding} H-mode, contrary to most other FPP concept regimes. Discarding the requirement to maintain $P_\mathrm{SOL}$ above the L-H power threshold grants freedom to use seeded impurities to radiate heat in the edge, lowering $\sub{P}{SOL}$ and thereby reducing divertor heat loads. Noble gas impurity seeding in PT L-mode experiments on DIII-D has shown enhanced confinement with low $P_\mathrm{SOL}$ \cite{McKee_impurity}, similarly to NT. Employing both NT shaping and seeded impurities allows more control over $\sub{P}{fus}$ and $\sub{P}{SOL}$ and potentially easier divertor integration than with NT alone. In a reactor, high-Z noble gas impurities such as krypton and xenon are expected to be most useful given their ionization temperatures, because they primarily cause radiation in the edge, decreasing $\sub{P}{SOL}$ significantly with minimal effect on $\sub{P}{fus}$. 

\begin{figure}
    \centering
    \includegraphics[width=0.5\textwidth]{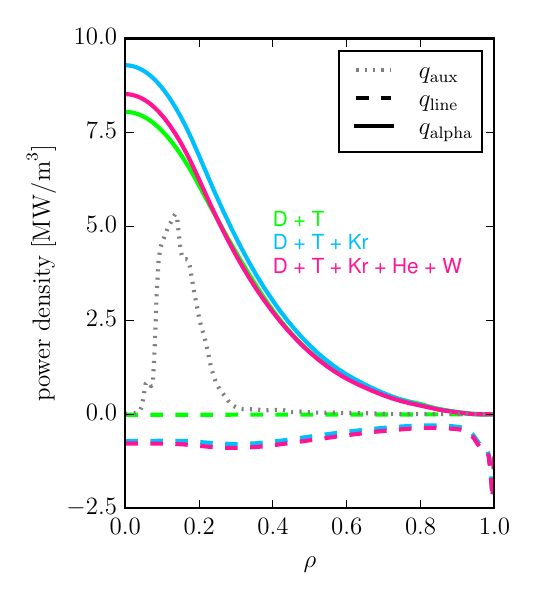}
    \caption{In the high-field case, power density profiles from auxiliary power, line radiation, and alpha power are plotted by the dotted, dashed, and solid lines respectively for three distinct fuel mixes. The D and T only mix is colored green, the D, T, and Kr mix is colored blue, and the D, T, Kr, He, and W mix is colored pink.}
    \label{fig:Prad_profiles}
\end{figure}

In all simulations mentioned thus far in this work, we included only krypton impurities at a fraction of 0.001 unless otherwise noted. Note that all impurity density profiles in this work are set to scale with the electron density profile, which is a limitation of this work. These results are likely to be affected by impurity transport causing impurity density profiles to differ significantly from the electron density profile. Tungsten is the leading candidate for plasma facing component in FPPs and so tungsten ions will likely also be in a reactor-class plasma along with helium ash, so its effect should be more closely studied in future work on the subject. 

To better assess the impacts of impurities on fusion performance in an NT FPP, we plot power density profiles from line radiation in the high-field base case with three distinct impurity combinations in figure \ref{fig:Prad_profiles}. In this figure, the power density profiles of the additional line radiation from including krypton (Kr) and tungsten (W) is plotted by the dashed lines in context of the alpha power density (solid lines) and auxiliary power density (dotted line). Note that the auxiliary power is the same for all three impurity combinations. The dilution effect of adding helium at a fraction of 0.02 in a D, T, and Kr mix results in a decrease in $\sub{P}{fus}$ of only 121 MW, or about 9\% of $\sub{P}{fus}$ in the D, T, and Kr only mix. The power density profiles of a D, T, Kr, and He mix are not shown in figure \ref{fig:Prad_profiles} because they overlap the D + T + Kr + He + W profiles. Including tungsten at a fraction of $1.5\times 10^{-5}$ does not significantly affect radiated power, and the $\sub{P}{fus}$ difference between the D, T, and Kr profile and the D, T, Kr, He, and W profile in figure \ref{fig:Prad_profiles} is only 127 MW, or $\sim10\%$. Note that the presence of tungsten will be inescapable in any reactor class device using tungsten as a plasma facing material (the leading metal candidate) and that it elicits serious concern for radiative collapse due to its high atomic number. For example, the tungsten fraction used in EU-DEMO modeling is $10^{-5}$ \cite{LUX2015} and in any reactor-class tokamak the upper limit on tungsten fraction is likely to be on the order of several $10^{-5}$ \cite{Putterich_2010}. H-mode has the benefit that ELMs flush impurities out of the core \cite{Dux2014,Fedorczak2015}, but the tungsten fraction limit is also set by maintaining $\sub{P}{SOL} > \sub{P}{L-H}$, a limit that NT does not need to adhere to. Additionally, ELMs can cause sputtering, leading to increased tungsten influx \cite{ELMsputtering}. While impurity transport in NT is an area of ongoing research, preliminary analysis on diverted DIII-D NT experiments suggests rapid impurity transport. This is indicated by hollow impurity profiles and lower $\sub{Z}{eff}$ in NT plasmas than PT plasmas with similar confinement factors \cite{Sciortino_2022}. 

\begin{figure}
    \centering
    \includegraphics[width=0.5\textwidth]{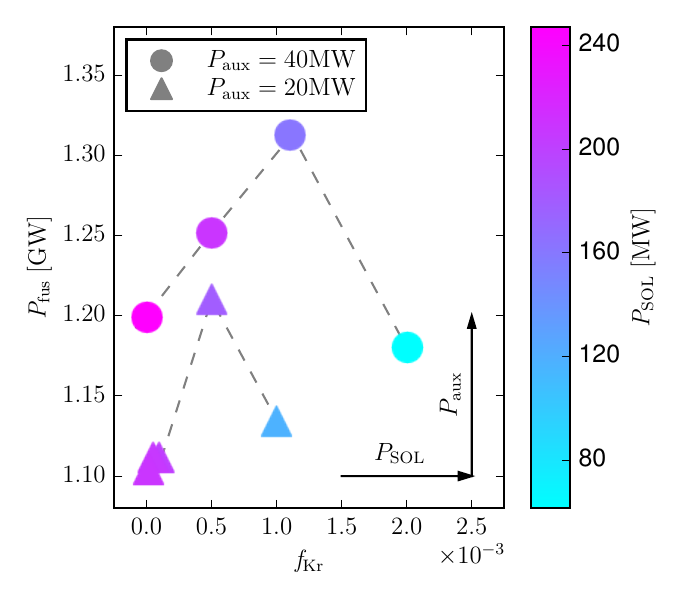}
    \caption{In the high-field case, the fusion power in mega-watts is plotted versus krypton fraction. Each point represents a converged simulation with only krypton impurity in a D-T fuel mix. The circular points are with 40 MW of input power and triangular points are with 20 MW of input power. Scrape-off layer power decreases with increasing impurity fraction for both $\sub{P}{aux} = 20$ MW and $\sub{P}{aux} = 40$ MW. Increasing $\sub{P}{aux}$ at a given impurity fraction increases fusion power.}
    \label{fig:imp_scan}
\end{figure}

In figure \ref{fig:imp_scan}, we plot $\sub{P}{fus}$ against krypton impurity fraction $\sub{f}{Kr}$ in the high-field case. The circles plotted are at 40 MW of auxiliary power while the triangles are at 20 MW of auxiliary power. The color bar gives $\sub{P}{SOL}$. Figure \ref{fig:imp_scan} shows that the same downward trend in $\sub{P}{SOL}$ versus $\sub{f}{Kr}$ change is seen at $\sub{P}{aux} = 40$ MW and when $\sub{P}{aux} =20$ MW. This is promising for prospective direct $\sub{P}{SOL}$ control using impurity seeding. However, figure \ref{fig:imp_scan} also shows that a small change in $\sub{f}{Kr}$ results in a large change in $\sub{P}{SOL}$, and research into whether this level of impurity control in a reactor will be possible is ongoing\cite{EldonSpecialIssue, kallenbach_impurity_2013}. Additionally, figure \ref{fig:imp_scan} shows a peaked trend in $\sub{P}{fus}$, indicating potential for $\sub{P}{fus}$ optimization via impurity fraction. This improvement in $\sub{P}{fus}$ from certain levels of additional impurity seeding while simultaneously decreasing $\sub{P}{SOL}$ has been observed in experiment in NT on DIII-D \cite{CasaliSpecialIssue} and encourages further study of this phenomenon for implementation in reactor modeling scenarios. For a given $\sub{f}{Kr}$, we also see that $\sub{P}{fus}$ is higher when $\sub{P}{aux} = 40$ MW than when $\sub{P}{aux}=20$ MW, indicating potential for $\sub{P}{fus}$ control via input power. At each auxiliary power, $\sub{H}{98y2}$ remained approximately constant over the scan of $\sub{f}{Kr}$, indicating that improvement in confinement from impurities may overcome the decrease in $\sub{P}{fus}$ from dilution.

While the inclusion of radiative impurities could lead to benefits in the core performance and in the power handling properties of NT FPPs, we note that we do not include in this work any decrease in the edge temperature resulting from significant radiation just inside of the separatrix. The impurities studied above were chosen to consolidate radiation in the core region \cite{kallenbach_impurity_2013}, but any drop in the temperature boundary condition will lead to a decrease in fusion power, as discussed in section~\ref{sec:edge_characterization}. As such, proper characterization of the role of impurities with a full impurity transport code that extends out into the SOL is needed for more robust NT FPP design, and should be the subject of future work.

\section{Conclusion}

\label{sec:conclusion}
These studies demonstrate the feasibility and flexibility of the NT reactor concept. Even with edge pressure conditions low enough to be infinite-$n$ ballooning stable, a variety of operating points exist in which FPP-relevant fusion power is possible ($\sim400-500$ MW for a MANTA-like device with 40 MW input power \cite{MANTA}), with opportunities for increased fusion gain by decreasing input power, increasing fueling, or optimizing seeded impurity fraction. Due to the lack of a requirement to maintain H-mode, NT also grants the freedom to increase the seeded impurity fraction. This can increase radiation in the edge to bring scrape-off layer power down to acceptable levels for simple divertor integration ($<40$ MW for a MANTA-like device with a separatrix density of 0.9 $\times 10^{20}$/m$^3$ \cite{MANTA}). 

The performance of an NT reactor will be heavily dependent on the edge condition. This dependency is not unique to NT but is perhaps of greater consequence in NT designs than in PT designs due to the present limitations in modeling the NT edge. Changing the edge temperature by 1 keV was found to have a more profound effect on the fusion power density than changing the major radius by 2 meters in a high-field case, though we note that this is without adjusting scans to account for changes in global stabilization parameters such as $\sub{q}{95}$. Additionally, we found the temperature at $\rho = 0.8$ to have a significantly larger effect on fusion power than auxiliary heating in both the high-field and high-volume cases, though the high-field case exhibited stronger dependence than the high-volume case. We reiterate that the density and temperature chosen for scans over other parameters in this work were to satisfy $\sub{H}{98y2} \approx 1$ and $\sub{f}{Gr} \approx 1$ while meeting temperature convergence in TGYRO with density at the Angioni peaking. We have not explored the extent to which these edge parameters can be accessed, and rely heavily on experimental observations to inform this work.

Though the effect of triangularity on fusion power density is minimal, fusion power density is found to consistently increase with less negative triangularity, suggesting that the optimal triangularity for an NT FPP may be that which is just negative enough to sustain ELM-free operation. Targeting this minimum negative triangularity will also have the added benefit of reducing vertical instability concerns \cite{Guizzo2024}. In this work, the core was modeled from $\rho = 0$ to $\rho = 0.8$, so future work should prioritize high fidelity transport modeling in the edge where the triangularity is strongest to identify any benefits attributable directly to the negative triangularity geometry, like that seen in reference \cite{DiGiannatale_2024}. 

For a high-field case with $\sub{f}{Gr} \approx 1$ and $\sub{H}{98y2} \approx 1$, we demonstrated a possible model for extrapolation to $\rho = 1.0$ that relates pedestal width to height based on the stability found from infinite-$n$ ballooning models. Thus for a given edge pressure boundary condition, there are a variety of pedestal shapes that remain ballooning stable. No scans of the high-volume case exceeded the infinite-$n$ ballooning stability boundary, so an increase in $\sub{R}{maj}$ could enable an increase in fusion power on two fronts: from the increased volume as well as the increased pressure boundary condition \cite{Nelson_2022}. 

The main advantages of NT come from the improved pressure edge condition over PT L-mode supported by experiment and the absence of a requirement to remain in H-mode allowing flexibility in impurity and heating requirements. We found that the temperature boundary condition and auxiliary power needed to reach fusion powers between 500-900 MW is below or on the order of representative PT H-mode FPP concepts. Future work to develop a physics-based model for the behavior of an NT reactor edge must be prioritized to more accurately predict fusion performance, though we have shown in this work that feasible core operation can be attained at a variety of edge conditions. To further evaluate the feasibility of integrating a given operating point with simple divertor operation, whole-device modeling like that done in reference \cite{MANTA} could illuminate the engineering trade-offs attributable to the potential plasma core trade-offs described here, such as pressure boundary condition, impurity content, geometry, and heating. Ultimately, there are likely a variety of core operating points for a NT tokamak FPP with competitive fusion power to leading PT H-mode concepts even at relatively high radiation, at low input power, and with a ballooning-stable pressure boundary and significantly reduced scrape-off layer power. 

\section{Acknowledgements}
Most calculations performed in this study were completed through the OMFIT framework \cite{MeneghiniOMFIT}, and the generated equilibria are available upon request. This work was supported by the U.S. Department of Energy, Office of Science, Office of Fusion Energy Sciences under Award DE-SC0022272.

\section{References}

\printbibliography[heading=none]

\appendix
\section{Full TGYRO Convergence}
\label{apx:convergence}
We define ``full" convergence to be met when the residual between the total flux calculated in the TGYRO/TGLF model ($\sub{f}{tot}$) and the target flux from power balance ($\sub{f}{tar}$) for each point between $\rho = 0.35$ and $\rho = 0.8$ is less than or equal to 0.02. Here the residual is defined as 
\begin{equation}
    \frac{(\sub{f}{tot}-\sub{f}{tar})^2}{\sub{f}{tot}^2+\sub{f}{tar}^2}.
\end{equation}
Converging from a residual of 0.02 to a residual of 0.00 resulted in less than a 5\% change in $\sub{P}{fus}$ in a few representative cases, so pushing convergence past this point is not expected to significantly change the qualitative results presented in this work. Flux matching between $\rho = 0.0$ and $\rho = 0.35$ is challenging but has ultimately been shown to have a marginal effect on output fusion power due in part to the relatively small plasma volume in the core compared to the edge \cite{Rodriguez-Fernandez_2024, Mariani_2024}.
\end{document}